\documentclass[aps,prd]{revtex4-2}
\usepackage{amsmath}
\usepackage{ulem}
\usepackage{color}
\usepackage{graphicx}
\usepackage{epsfig}
\usepackage{subfig}
\usepackage{bm}
\usepackage{hyperref}
\usepackage{natbib}
\usepackage[justification=raggedright,singlelinecheck=false]{caption}
\usepackage{subcaption}
\usepackage{booktabs}
\usepackage{siunitx}
\def\be {\begin{equation}}
\def\ee {\end{equation}}

\def\bea {\begin{eqnarray}}
\def\eea {\end{eqnarray}}

\begin{document}
	%\linenumbers
	
	\title{Numerically studying Pesticide diffusion in air using Langevin formalism}        
         \author{{Utkarsh Patel}}
	\author{{Sabyasachi Ghosh}}
         \author{{Prasanta Murmu}}
         \email{utkarshp@iitbhilai.ac.in}
	\email{sabya@iitbhilai.ac.in}
        \email{prasantamurmu@iitbhilai.ac.in}
	\affiliation{Department of Physics, Indian Institute of Technology Bhilai, Kutelabhata-491002, India}
	\begin{abstract}
		      The use of pesticides for enhancing crop yield and preventing infestations is a widespread agricultural practice. However, in recent years, there has been a growing shift toward traditional chemical-free organic farming. Regulatory frameworks impose specific distance requirements between organic farms and neighboring lands where chemical pesticides are used to minimize cross-contamination. In this work, we numerically analyze the spread of pesticide droplets to adjacent fields under varying weather conditions, providing a systematic analysis that highlights conditions where existing guidelines might require reassessment. We employ the formalism of the Langevin equations to model the diffusion of pesticide particles and their transport due to wind and other environmental factors. Assuming a non-relativistic, classical diffusion framework, we track the dispersion of commonly used pesticides to assess their potential contamination range. We present our key findings, discuss their implications, and, toward the end, outline possible directions for future research. 
	\end{abstract}
	
	%\pacs{12.38.Mh,25.75.-q,24.85.+p,25.75.Nq}
	\maketitle
\section{Introduction}
\label{sec:intro}
During pesticide application, a fraction of the dosage is lost to the atmosphere. This application loss refers to the total fraction of the applied dosage that does not reach the intended target area. A significant part of this loss occurs when droplets move to non-target surfaces through the air, a process known as spray drift. While pesticide is applied by aerial spraying, a substantial fraction may drift into the atmosphere or reach the air after application due to evaporation or volatilization from soils and crops~\cite{toxics11100858}. It can also be transported downwind, as influenced by wind direction and speed. The atmosphere, being a complex medium, enables the long-distance transport of pesticides and serves as a critical medium for the persistence of pesticides in solid, liquid, and gaseous forms. Persistent organic pollutants (POPs) can linger in the environment and spread globally due to their physicochemical properties, posing a concern for both human health and wildlife. It has been found that 16 out of the 24 officially recognized POPs under the Stockholm Convention are pesticides.

During ground or aerial spraying, fine droplets increase the potential for spray drift to unintended areas, leaving a fraction of the pesticide immediately airborne. The lifetime of pesticides in the air depends on their gas-phase reactivity, while sorption onto aerosol particles significantly extends their half-lives in the atmosphere~\cite{Socorro2016}. The application method, including tools and techniques, influences the rate and extent of spray particle drift in addition to primary factors such as pesticide formulation, environmental conditions, soil properties, and crop response~\cite{vandenBerg1999}. The dominant factors governing pesticide droplet emission into the air during application can be broadly categorized as technical and environmental. Reference~\cite{Hofman2001} grouped them into the following: (a) spray characteristics, such as volatility and viscosity of the pesticide formulation, (b) equipment and application technique, (c) weather conditions at the time of application (wind speed and direction, temperature, relative humidity, and atmospheric stability), and (d) operator care, attitude, and skill~\cite{GIL20072945}. However, spray drift is an inherent component of every spraying operation. The challenge with spraying small insecticide droplets is ensuring their deposition on the target, as they tend to remain airborne and drift long distances due to their low weight and small size.

Effective coverage is crucial when applying insecticides and fungicides due to the extremely small size of target organisms. Small to medium-sized droplets are generally preferred as they tend to provide more thorough and uniform coverage. The second most significant factor after wind speed and direction is the size of the pesticide droplet, which influences both coverage and drift. It is therefore wise to optimize droplet size during pesticide application, as coverage typically reduces with increasing droplet size. Systemic herbicides are often effective when applied with larger droplets, whereas contact-type insecticides and fungicides generally require smaller to medium-sized droplets to ensure better coverage and more effective results~\cite{Hofman2001}.

Diffusion equations describe the spatial random movement of pesticide particles and are often modeled using Fick’s or the Fokker-Planck diffusion laws. In contrast to Fick’s law, the Fokker-Planck equation~\cite{Bengfort} accommodates spatial heterogeneities and spatiotemporal pattern formation. Although many models simplify diffusion to be spatially homogeneous, this assumption can render Fick’s law and the Fokker-Planck equation superficially equivalent~\cite{Fick1855,Fick1858,Fokker,Planck1917}. However, when spatial inhomogeneities are considered, these two diffusion formalisms yield different outcomes~\cite{Bengfort}.

Previous studies on spray dynamics have explored factors like initial droplet velocity, size, and subsequent dispersion behavior. Dorr et al.~\cite{Dorr2013} observed that velocity is highest near the nozzle and decreases with distance, while smaller droplets decelerate faster and reach terminal velocity sooner than larger ones. Research on droplet velocity is comparatively limited due to measurement challenges \cite{Nuyttens,Dorr2013}. Several studies emphasize the correlation between droplet size and spray drift potential, noting that finer droplets ($<150 \mu$m) are more prone to evaporation and atmospheric transport, especially those under $50\mu$m, which are difficult to control \cite{van1999}. Formulation type, ventilation, and time since application have also been shown to influence airborne pesticide concentrations, with aerosol and emulsifiable forms generally exhibiting higher initial levels than microencapsulated ones~\cite{Koehler1995}. Similarly, HCB has been detected at relatively uniform levels across large oceanic regions, indicating long-range atmospheric transport~\cite{WU2014477}. We have investigated the airborne dispersion tendencies and patterns for two commonly used pesticides: chlorpyrifos and hexachlorobenzene (HCB). Chlorpyrifos, an organophosphate pesticide, is a white crystalline solid with a strong odor, widely used for soil insect control via foliar application. However, several studies~\cite{coulston1972,Koehler,warner1980results} report long-term hazardous effects of airborne chlorpyrifos on humans and animals. Prolonged exposure can lead to toxic effects via skin absorption~\cite{Schiott1970}. Inhalation exposure in recently sprayed areas can affect the nervous system and lead to vision loss, muscle weakness, coma, or even death under high exposure. Hexachlorobenzene (HCB), a persistent organic pollutant, can become airborne and enter the human body via inhalation or through contaminated food, water, and soil, potentially causing liver damage and carcinogenic effects~\cite{IARC1979Supplement1,IARC1979Vol20,Cabral1986}. Besides its use as a fungicide, HCB is a by-product of several industrial processes and is often found as an impurity in pesticide formulations. The U.S. EPA classifies HCB as a probable human carcinogen (Group B2), and it is one of the 12 POPs listed under the Stockholm Convention.

The aim of our study is to numerically simulate the airborne dispersion of chlorpyrifos and HCB at varying initial droplet sizes, fixed spray densities, and under different wind speeds. We analyze the resulting tangential and radial displacement dynamics over time. This pattern provides insight into the potential aerial transport and drift of pesticide sprays, helping predict deposition on both target and non-target surfaces, and contributing to the understanding of their environmental impact in agricultural use. The structure of the paper is as follows. In Section~\ref{sec:intro}, we introduce the context and motivation of our study. Section~\ref{sec:formalism} presents the Langevin dynamics-based mathematical formalism used for simulating pesticide droplet dispersion. Section~\ref{sec:res} provides the simulation results and interpretation. Section~\ref{sec:conc} summarizes our findings and proposes directions for future research. An appendix contains detailed derivations supporting our analysis.

\section{Formalism: Solving the Langevin Equation Numerically via Monte Carlo Techniques}
\label{sec:formalism}

To model the motion of pesticide droplets in the air, we employ the Langevin equations~\cite{reif1965fundamentals}, which describe the dynamics of droplets of mass \( M \) under the influence of gravitational and stochastic forces. The mathematical form of these equations is expressed as: 

\begin{eqnarray}
    \frac{dr_i}{dt} &=& v_i, \label{eq:langevin1} \\
    M \frac{dv_i}{dt} &=& -\lambda v_i + \xi(t) + F_g. \label{eq:langevin2}
\end{eqnarray}

where \( dr_i \) and \( dv_i \) represent the changes in position and velocity over a discrete time step \( dt \), respectively. The subscript \( i \) denotes the Cartesian components of the position and velocity vectors. The parameter \( \lambda \) in Eq.~(\ref{eq:langevin2}) is the drag coefficient.

The first term on the right-hand side of Eq.~(\ref{eq:langevin2}) accounts for the dissipative force due to air resistance, while the second term represents the diffusive (stochastic) force, where \( \xi(t) \) follows a diffusion coefficient \( D \). The noise term \( \xi(t) \) is modeled as white noise~(as employed in Ref.~\cite{Das2020}), satisfying \( \langle \xi(t) \rangle = 0 \) and \( \langle \xi(t) \xi(t') \rangle = D \delta(t - t') \), where the correlations are instantaneous, following a Dirac delta function. The third term, \( F_g \), represents the gravitational force acting on the droplet, given by \( F_g = Mg \), where \( g = 9.8 \) m/s\(^2\).

To account for environmental wind effects, we may assume a constant horizontal wind velocity \( u(t) \), which remains uniform across the field but may vary over time:

\begin{equation}
    u(t) = u_0 + \Delta u(t),
\end{equation}

where \( u_0 \) is the mean wind velocity, and \( \Delta u(t) \) represents fluctuations due to atmospheric turbulence, which can be modeled as a stochastic component. Since wind speed over agricultural fields is influenced by diurnal cycles and seasonal variations, this approach captures a more realistic representation of pesticide drift.
With the inclusion of wind effect, Eq.~\ref{eq:langevin1} is modified as:
\begin{equation}
\frac{dr_i}{dt} = v_i+u(t)
\label{eq:langevin3}
\end{equation}

The Langevin equations (Eqs.~(\ref{eq:langevin1}) and~(\ref{eq:langevin2})) are solved numerically using Monte Carlo techniques. The initial spatial position of a droplet is taken as \( (x, y, z) = (0, 0, H_0) \), where \( H_0 \) is the initial release height, corresponding to the height at which pesticides are sprayed over crops. The initial spray velocity is distributed randomly within the \( x \)-\( y \) plane, while its \( z \)-component points downward and depends on the spray angle relative to the horizontal. The gravitational force acts along the downward \( z \)-axis. 

We consider a range of droplet radii \( R \) from 30~\(\mu\)m to 100~\(\mu\)m and initial ejection velocities \( V_0 \) from 1~m/s to 5~m/s. Droplet sizes below 30~\(\mu\)m are not considered, as they tend to evaporate rapidly before experiencing any significant horizontal descent~\cite{van1999}. The mass of a droplet is estimated using:
\begin{equation}
    M = \frac{4}{3} \pi R^3 \rho,
    \label{eq:mass}
\end{equation}

where \( \rho \) is the density of the pesticide droplet. This formulation allows us to track the dispersion of pesticide particles under different wind conditions and evaluate their potential drift to nearby fields.

\section{Result and discussions}
\label{sec:res}
Based on the above mathematical framework, the most significant parameters that contribute to the process of pesticide transportation from spray nozzle to plant are presented below:  

\begin{enumerate}
	\item Concentration of the pesticide in the solution.
	\item Speed, height, and angle of the spray.	
	\item Physical and chemical characteristics of the droplet of pesticide solution, such as temperature ($T_D$), volume, mass, and chemical reagent present in it.	
	\item Properties of air around the agricultural field, such as humidity, temperature ($T$), pressure, and wind speed.
\end{enumerate}

A 5\% mL/g concentration of the pesticide solution in water is adopted in this study, as it aligns with commonly employed concentrations in real-world applications, thereby justifying its use as a baseline assumption~\cite{Teske2002AgDRIFT}. For the simulation, we assume that the sprayer is operated from an average person's height, taken to be $1.7$~m. The angle of spray, for simplicity, is chosen to be aligned at $30^\circ$ below the horizontal plane, implying that a small component of initial spray velocity from the nozzle helps the droplet with their vertical descent. The horizontal component of the spray velocity is considered to be random and is parameterized by an angle that is assigned a random value in each simulation run. As a result, the average initial velocity in the horizontal direction is calculated by $v_{\text{hor}} = \sqrt{v_x^2 + v_y^2}$ for each run.

\begin{table}[h!]
\centering
\begin{tabular}{@{}lcc@{}}
\toprule
\textbf{Property (units)} & \textbf{Chlorpyrifos} & \textbf{HCB} \\
\midrule
Molecular Weight (g/mol)     & 350.59~\cite{Merck1989}         & 284.78~\cite{NTP_Hexachlorobenzene}       \\
Density (\si{g/cm^3})         & 1.40~\cite{Verschueren1983}           & 2.04~\cite{NTP_Hexachlorobenzene}         \\
Vapor Pressure (mmHg at 25\,°C) ~~~~& 1.87 × 10\(^{-5}\)~\cite{Merck1989} ~~~~& 3.0 × 10\(^{-5}\)~\cite{EXTOXNET_HCB} \\
\bottomrule
\end{tabular}
\caption{Comparison of selected physicochemical properties of Chlorpyrifos and Hexachlorobenzene (HCB).}
\label{tab:1}
\end{table}

For this study, we consider two different pesticide chemicals: \textit{Chlorpyrifos} and \textit{Hexachlorobenzene} (HCB). Few chemical properties of these chemicals are listed in Table ~\ref{tab:1}. The effective density of the pesticide solution is calculated separately for each chemical. From Table~\ref{tab:1}, it can be referred that the density of Chlorpyrifos is $1.40\text{ gm}/\text{cm}^3$ and the density of HCB is $2.04\text{ gm}/\text{cm}^3$. For a $5\%$ mL/g solution of these chemicals in water at $30\,^{\circ}\text{C}$, we calculate the effective solution density. The details of this calculation are straightforward and can be referred from the appendix~\ref{app:A}. The effective solution density for Chloropyriphos at 5\% concentration and 30\,°C is approximately $1036.13 \text{ kg/m}^3$. Similarly, we may perform a calculation for the effective density of HCB, after which the effective solution density for HCB at 5\% concentration and 30\,°C comes out to be approximately $1048.92 \text{ kg/m}^3$. The computed effective density for each chemical is used to determine the mass of a droplet for a fixed radius (assuming spherical droplets) using the relation given in Eq.~\ref{eq:mass}.

\begin{figure}[h!]
  \centering
  \includegraphics[width=0.5\textwidth]{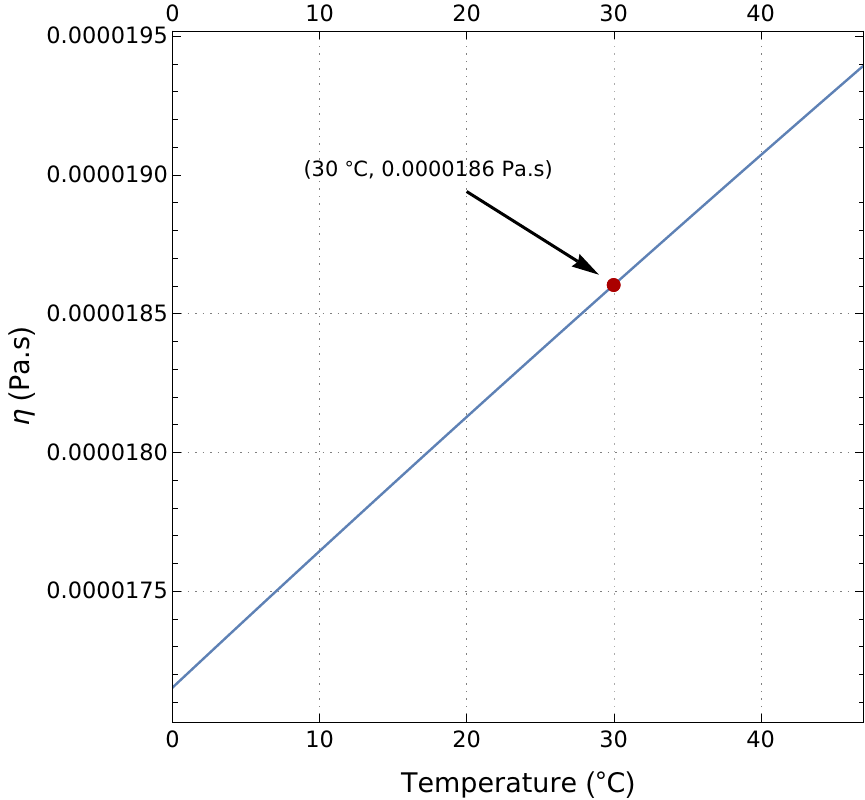}
  \caption{\small Temperature dependence of the dynamic viscosity of air, based on Equation~\eqref{eq:air_viscosity}.}
  \label{fig:visc}
\end{figure}

Additionally, a temperature-dependent viscosity model~\cite{white2006viscous} is adopted to account for the effects of ambient air temperature on the droplet dynamics, and is expressed as:
\begin{equation}
\eta~(T) = 1.716 \times 10^{-5} \left( \frac{T}{273.15} \right)^{3/2} \frac{273.15 + C_s}{T + C_s} \ \text{Pa}\cdot\text{s},
\label{eq:air_viscosity}
\end{equation}

where, \( \eta(T) \) is the dynamic viscosity of air at temperature \( T \) (in K) and $C_s = 111 \ \text{K}$ is the Sutherland's constant~\cite{sutherland1893viscosity}. This temperature dependence of air viscosity is graphically shown in Figure~\ref{fig:visc}. Using Eq.~\ref{eq:air_viscosity}, $\eta$ for air at 303 K~(30 °C) is equal to $1.859\times10^{-5}$ Pa.s.

Correspondingly, the linear drag coefficient and the diffusion coefficient are also calculated for the droplet in each case based on the physical properties of the respective chemical solutions. The linear drag coefficient \( \lambda \), which characterizes the resistance experienced by a spherical droplet moving through a viscous fluid, can be expressed using Stokes' law. For a droplet of radius \( R \) in a medium with dynamic viscosity \( \eta \), the drag coefficient is given by~\cite{stokes1851}:
\begin{equation}
\lambda = 6 \pi \eta R
\label{eq:drag_coefficient}
\end{equation}
This linear form of drag is valid under conditions of low Reynolds number, where viscous forces dominate over inertial forces, a typical scenario for microscopic droplets in air or water.

The diffusion coefficient \( D \) associated with the random thermal motion of the droplet is related to the drag coefficient through the Einstein relation in momentum space as~\cite{beale1996statistical}:
\begin{equation}
D = \frac{K_B T}{\lambda}
\label{eq:einstein_relation}
\end{equation}
where \( K_B = 1.38 \times 10^{-23} \, \text{J/K} \) is the Boltzmann constant, and \( T \) is the absolute temperature in Kelvin. This relation reflects the balance between thermal energy and viscous damping, encapsulating the fluctuation-dissipation principle. With all these information about the system, we are ready to investigate the results of numerically solving the above equations for various different input parameter values and conditions.

\begin{figure}[h]
    \centering
    \begin{minipage}{0.48\textwidth}
        \centering
        \includegraphics[width=\textwidth]{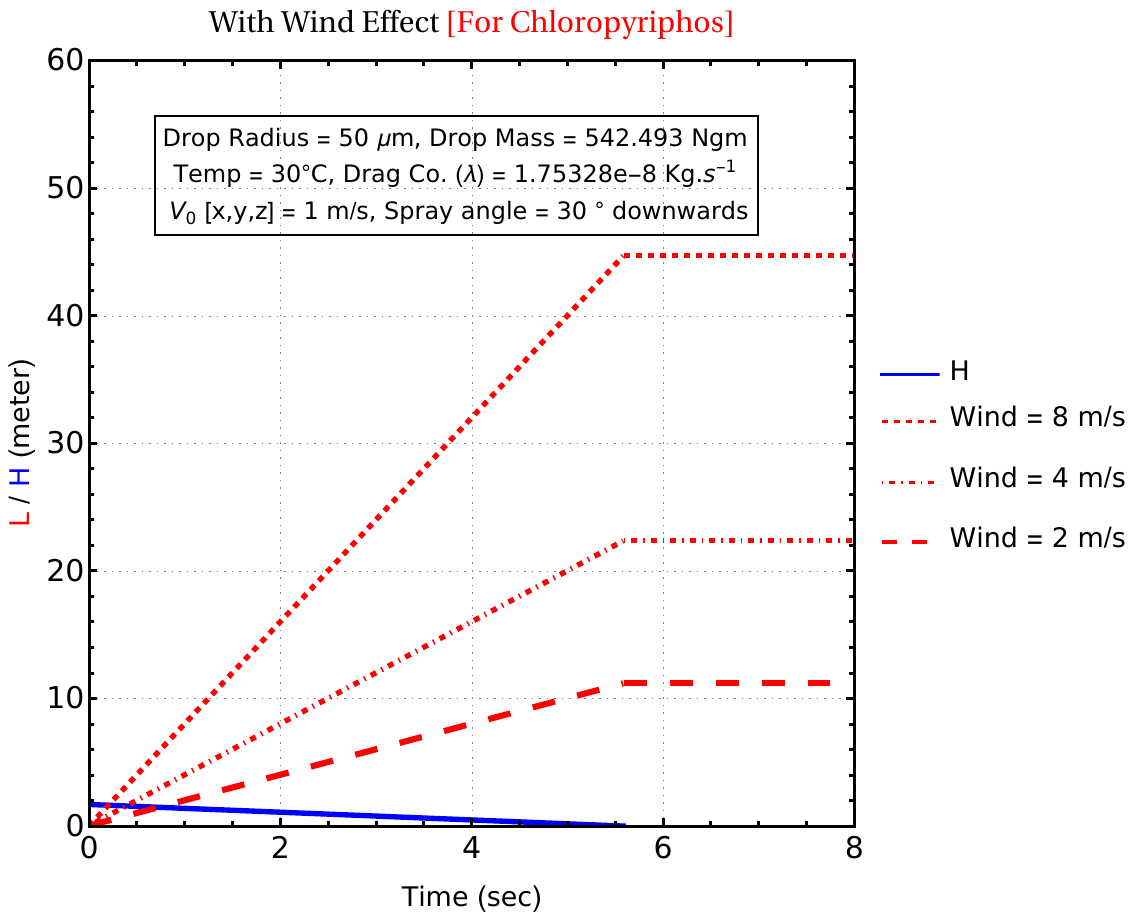}
        %\caption{\textbf{Chlorpyrifos}}
        \label{fig:chlorpyrifos}
    \end{minipage}
    \hfill
    \begin{minipage}{0.48\textwidth}
        \centering
        \includegraphics[width=\textwidth]{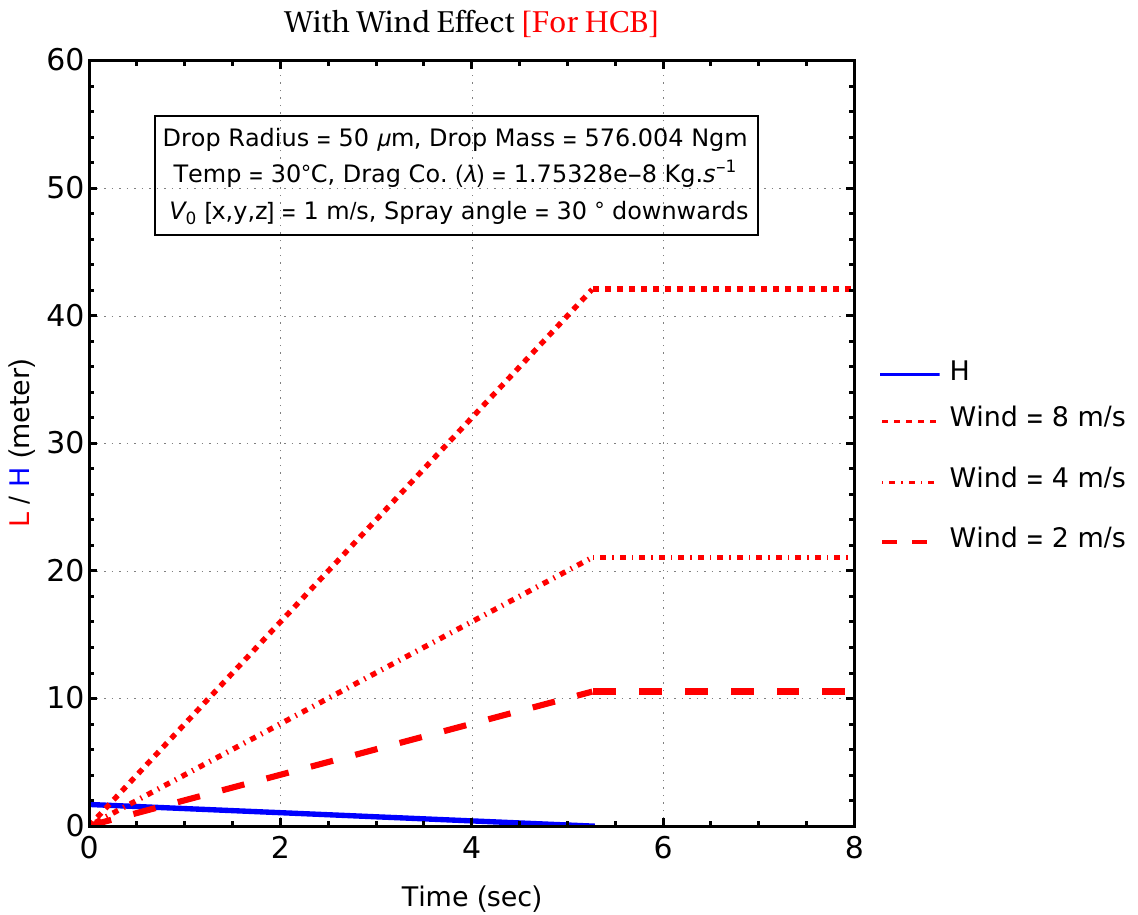}
        %\caption{\textbf{HBC}}
        \label{fig:hbc}
    \end{minipage}
    \caption{\small Numerical simulations of pesticide droplet dispersion under wind effects for \textbf{Chlorpyrifos (left panel)} and \textbf{HCB (right panel)}. The plots track both the height~($H$) and radial distance~($L$) of the droplets over time. The solid blue line represents the height ($H$), while the red lines indicate the radial displacement ($L$) for different wind speeds:  
    dashed line for 2 m/s, dot-dashed line for 4 m/s, and dotted line for 8 m/s. The simulations consider a droplet radius of 50 $\mu$m and a mass of 542.493 Ngm at a temperature of 30°C. The drag coefficient ($\lambda$) is taken as $1.75328\times 10^{-8}$ kg$\cdot$s$^{-1}$, spray angle is fixed at $30^\circ$ downwards from the horizontal and, the spray velocity of the droplet is taken to be $V_0 [x,y,z] = 1$ m/s.}
    \label{fig:wind_effect_pesticides}
\end{figure}

\begin{figure}[h]
    \centering
    \begin{minipage}{0.48\textwidth}
        \centering
        \includegraphics[width=\textwidth]{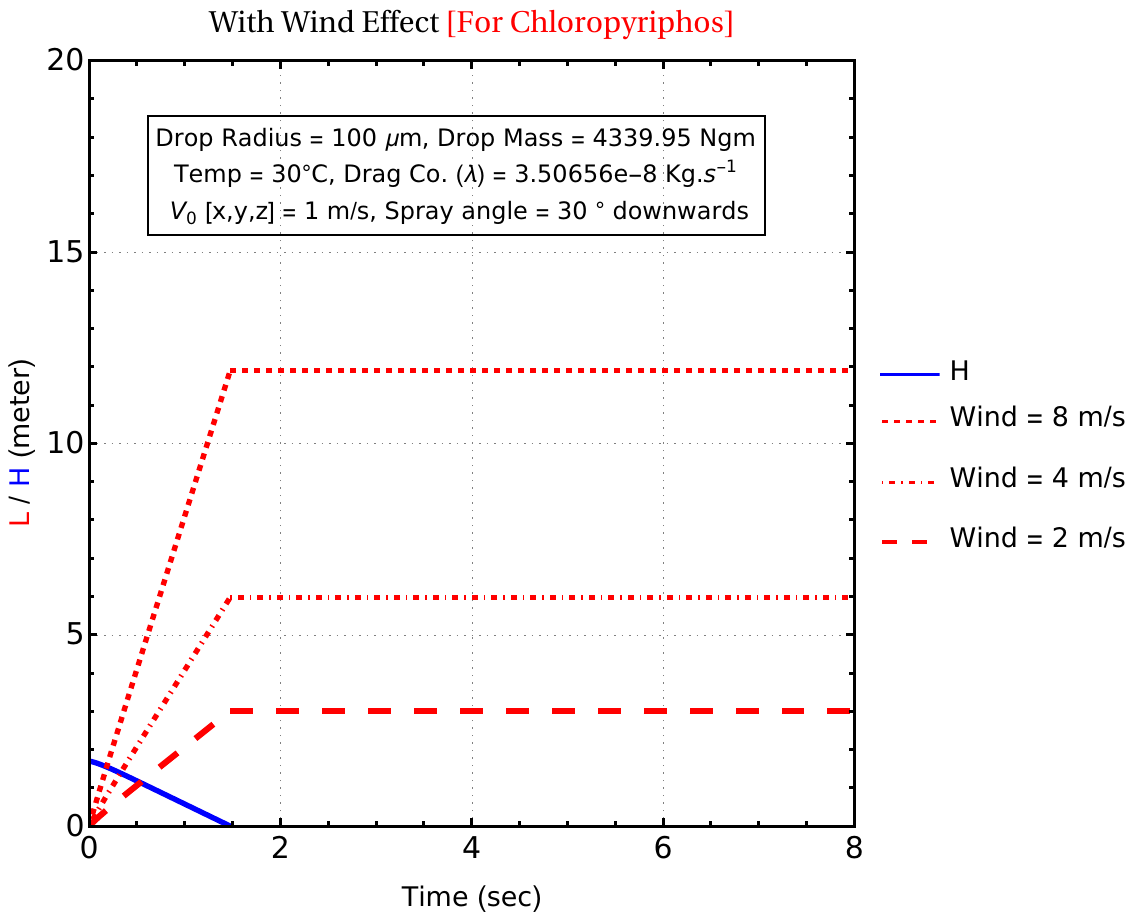}
        %\caption{\textbf{Chlorpyrifos}}
        \label{fig:chlorpyrifos2}
    \end{minipage}
    \hfill
    \begin{minipage}{0.48\textwidth}
        \centering
        \includegraphics[width=\textwidth]{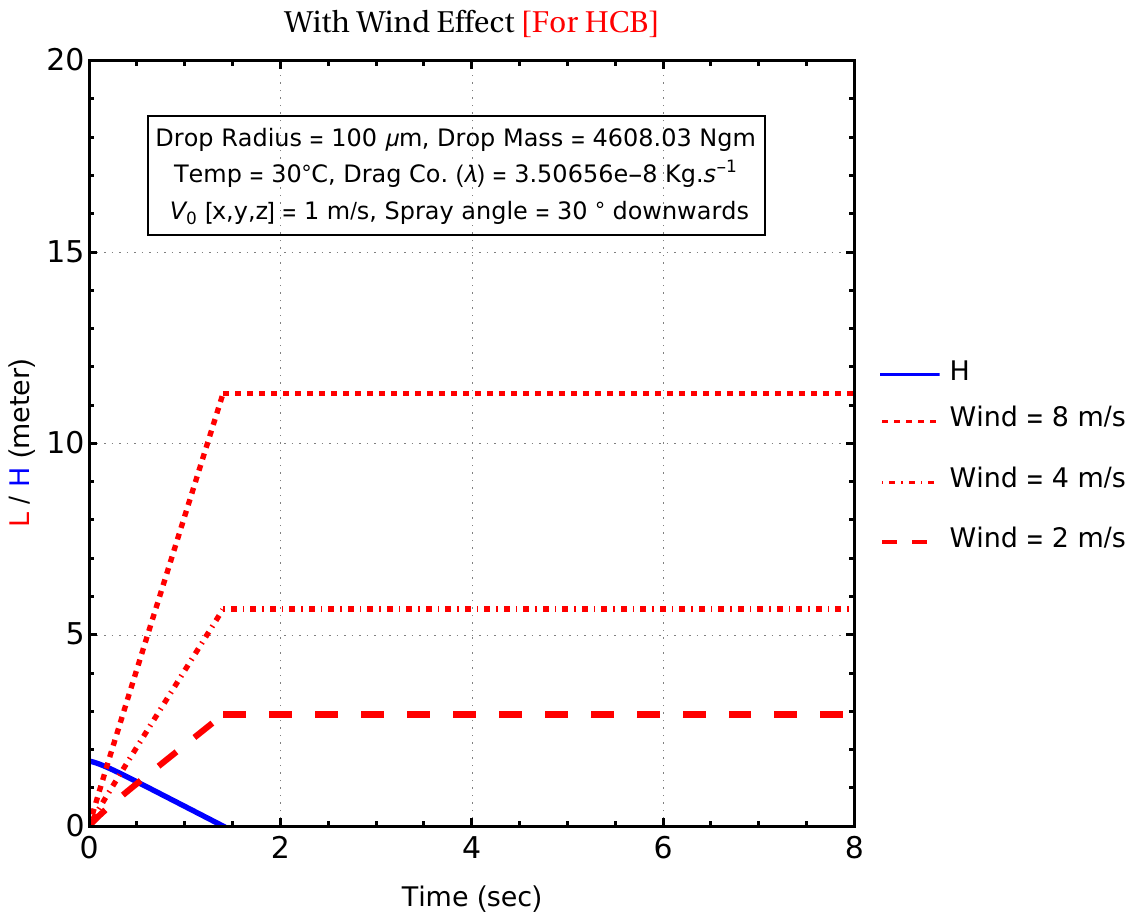}
        %\caption{\textbf{HBC}}
        \label{fig:hbc2}
    \end{minipage}
    \caption{\small Same as figure \ref{fig:wind_effect_pesticides}, but with input parameter, drop radius taken to be $100\mu$m here.}
    \label{fig:wind_effect_pesticides2}
\end{figure}

For our analysis, we have adopted the Monte Carlo simulation technique to calculate the final numerical results. Specifically, for each simulation, the initial velocity components and wind direction were randomized over a 1000 samples, and the results were averaged to obtain a statistically meaningful trajectory for the pesticide droplets. This methodology is consistently adopted throughout the analysis presented in this work.

\begin{figure}[h]
    \centering
    \begin{minipage}{0.48\textwidth}
        \centering
        \includegraphics[width=\textwidth]{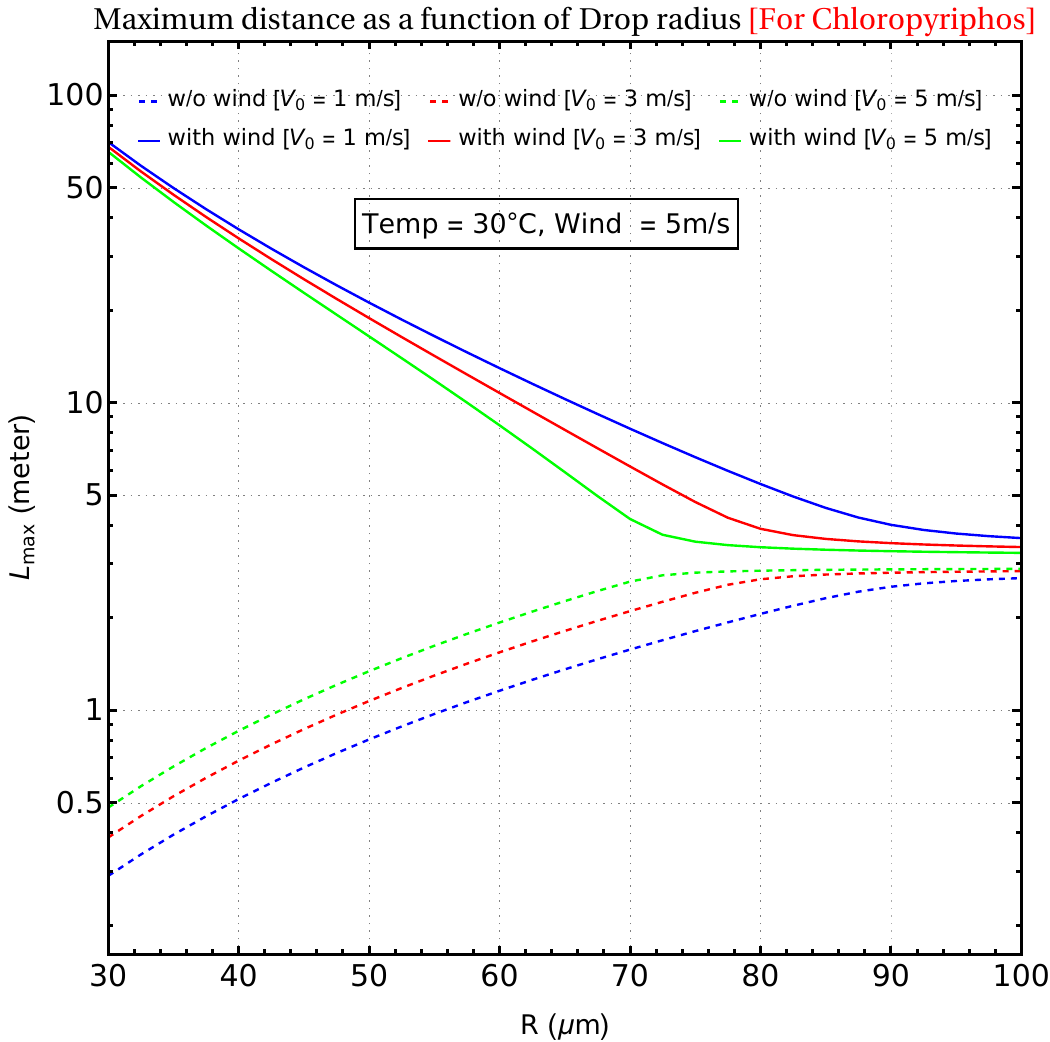}
        %\caption{Chlorpyrifos}
        \label{fig:chlorpyrifos_Lmax}
    \end{minipage}
    \hfill
    \begin{minipage}{0.48\textwidth}
        \centering
        \includegraphics[width=\textwidth]{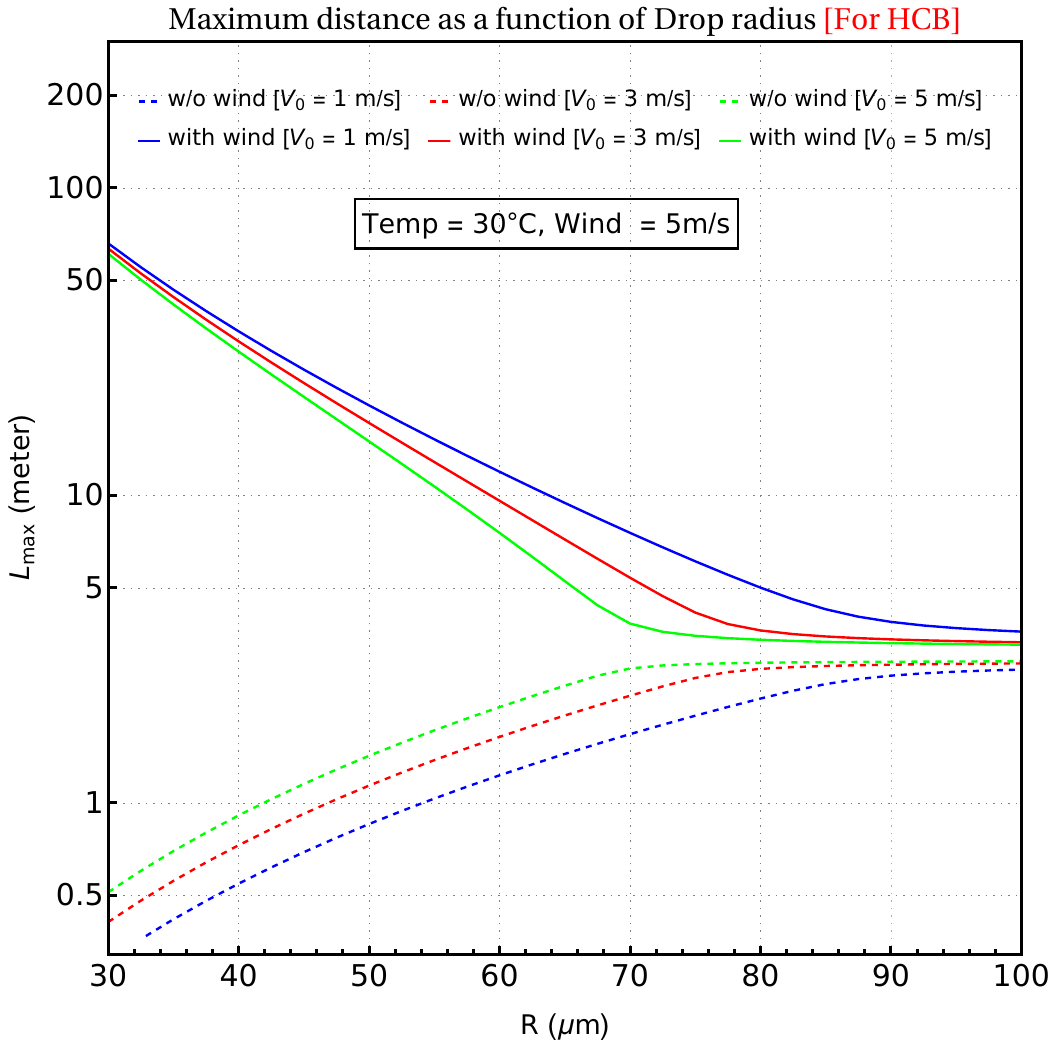}  % Change to actual filename for HCB
        %\caption{HCB}
        \label{fig:hcb_Lmax}
    \end{minipage}
    \caption{\small Maximum horizontal distance covered ($L_{\max}$) as a function of drop radius ($R$) for Chlorpyrifos \textbf{(left panel)} and HCB \textbf{(right panel)} for two different cases: (i) without wind, shown by dashed curves, and (ii) with wind effect, shown by solid curves. Each color represents a different initial velocity ($V_0$): Blue curves: $V_0 = 1$ m/s, Red curves: $V_0 = 3$ m/s, and Green curves: $V_0 = 5$ m/s. The simulations are performed at a temperature of 30°C, with ambient air viscosity of $1.859\times10^{-2}$ mPa$\cdot$s. The wind speed in the wind-included case is set to 5 m/s. This analysis demonstrates the significant influence of droplet size and wind effect on the maximum horizontal dispersion of pesticide droplets.}
    \label{fig:drop_radius_vs_Lmax}
\end{figure}

\begin{figure}[h]
    \centering
    \begin{minipage}{0.48\textwidth}
        \centering
        \includegraphics[width=\textwidth]{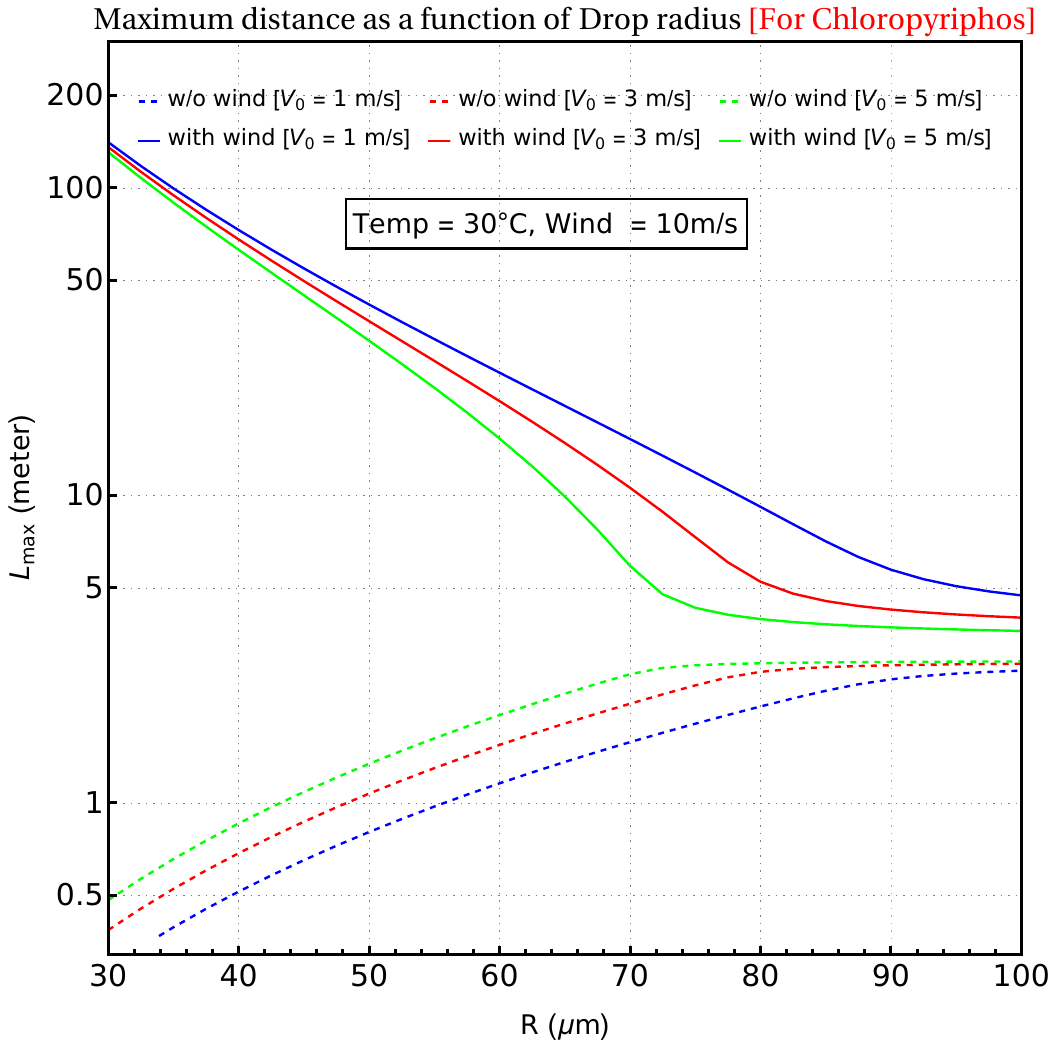}
        %\caption{Chlorpyrifos}
        \label{fig:chlorpyrifos_Lmax2}
    \end{minipage}
    \hfill
    \begin{minipage}{0.48\textwidth}
        \centering
        \includegraphics[width=\textwidth]{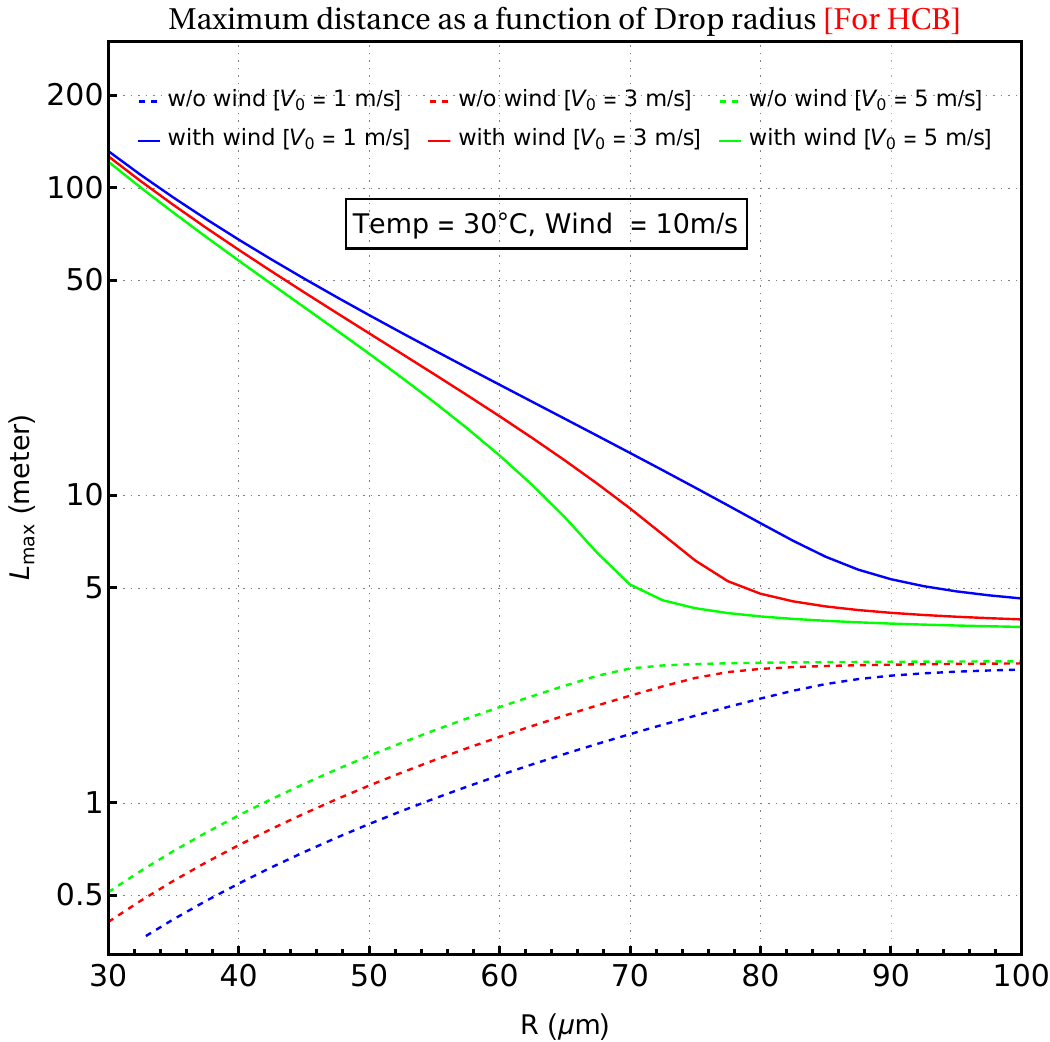}  % Change to actual filename for HCB
        %\caption{HCB}
        \label{fig:hcb_Lmax}
    \end{minipage}
    \caption{\small Same as figure \ref{fig:drop_radius_vs_Lmax}, but with input parameter, wind speed taken to be $10~\text{m/s}$ here.}
    \label{fig:drop_radius_vs_Lmax2}
\end{figure}

Figure~\ref{fig:wind_effect_pesticides} shows the solution of the Langevin equation for our system, where the total horizontal distance covered by the droplet ($L$) and the corresponding height ($H$) (in meters) from which the droplet is spread are plotted as a function of time (in seconds) for two different chemicals: \textit{Chlorpyrifos} (left panel) and \textit{HCB} (right panel). The blue solid line represents the droplet height ($H$), while the colored lines indicate the radial displacement ($L$) for different wind speeds: dashed for $2$~m/s, dot-dashed for $4$~m/s, and dotted for $8$~m/s. The input parameters for this simulation are as follows: the droplet radius is fixed at $50$~$\mu$m, with a mass of $542.493$~Ngm for Chlorpyrifos and $576.004$~Ngm for HCB. The surrounding air temperature is taken as $30^\circ$C, and the viscosity of the pesticide droplets is set to $0.89803$~mPa$\cdot$s. The drag coefficient is given by $\lambda = 1.75328\times 10^{-8}$ kg$\cdot$s$^{-1}$. The spray angle is fixed at $30^\circ$ downwards from the horizontal, and the initial spray velocity of the droplet is taken to be $V_0[x,y,z] = 1$~m/s.

From the figure, it can be observed that with an increase in wind strength, the maximum horizontal distance covered by a droplet of radius $50$~$\mu$m before falling to the ground increases significantly. However, the overall droplet dynamics remain largely unaffected by the chemical type, as seen by comparing the results from the left and right panels, which exhibit nearly identical behavior. This suggests that for the given input parameters, wind speed dominates the dispersion characteristics rather than the specific properties of the pesticide solution. These results provide an essential basis for further analysis, where we examine the impact of other environmental and chemical parameters on droplet dispersion and deposition patterns. To analyze the effect of drop radius on the maximum distance traveled as a function of time, in Figure~\ref{fig:wind_effect_pesticides2}, we check the same dependence, but for a droplet radius of $100,\mu$m. With the larger particle size (assuming a spherical drop), the mass of the droplet increases, thereby enhancing the influence of the gravity-dependent term in the Langevin equation. As a result, the droplet descends more rapidly, reducing its horizontal travel. Compared to the case of a $50,\mu$m droplet shown in Figure~\ref{fig:wind_effect_pesticides}, the larger droplet covers a shorter horizontal distance at all wind speeds. This trend is observed consistently for both Chlorpyrifos (left panel) and HCB (right panel) in Figure~\ref{fig:wind_effect_pesticides2}.

Figure~\ref{fig:drop_radius_vs_Lmax} illustrates the maximum horizontal distance covered ($L_{\text{max}}$) as a function of droplet radius ($R$) for both Chlorpyrifos (left panel) and HCB (right panel). The results are shown for two different cases: (i) without wind, represented by dashed curves, and (ii) with wind, represented by solid curves. Each color corresponds to a different initial velocity ($V_0$), with blue denoting $V_0 = 1$ m/s, red indicating $V_0 = 3$ m/s, and green corresponding to $V_0 = 5$ m/s. The input parameters for this analysis are: ambient temperature $T = 30^\circ$C, ambient air viscosity $\eta = 1.859\times10^{-2}$ mPa$\cdot$s, and wind speed $v_{\text{wind}} = 5$ m/s for the wind-included case. The results reveal that for droplet radii below approximately 60 $\mu$m, the droplets remain airborne for a significant period before settling. Consequently, if sufficient wind is present, such small droplets can be transported over long distances. Specifically, for radii below 40 $\mu$m, the droplets can travel well beyond $20$ meters before deposition, as evident from both the left and right panel plots. This phenomenon arises due to the interplay between gravitational and aerodynamic forces: for very small droplets, the gravitational force is weak compared to the drag force exerted by the surrounding air. As a result, after an initial transient phase, the droplet reaches a near-equilibrium state where the net vertical force approaches zero, leading to terminal velocity conditions. Under these circumstances, the horizontal drift due to wind becomes the dominant transport mechanism. These findings highlight the importance of droplet size in determining pesticide drift, emphasizing that for radii below a certain threshold, external factors like wind play a crucial role in determining the spatial distribution of the pesticide. For large droplet radii ($R \gtrsim 70~\mu\text{m}$), the maximum covered horizontal distance $L_{\text{max}}$ tend to saturate with increasing radius and eventually aligns with its value when no wind effects are taken into account. In this regime, gravitational settling dominates over drag and wind effects due to the higher mass and inertia of the droplets. As a result, these droplets fall more vertically with limited lateral drift, and the influence of initial velocity or wind becomes less significant. This leads to a plateau-like behavior in $L_{\text{max}}$ for larger droplets.

In Figure~\ref{fig:drop_radius_vs_Lmax2}, we plot the maximum horizontal distance ($L_{\text{max}}$) as a function of droplet radius ($R$) for both Chlorpyrifos (left panel) and HCB (right panel), considering a higher wind speed of $10$~m/s. With the wind speed tripled compared to the case in Figure~\ref{fig:drop_radius_vs_Lmax}, a corresponding increase in the horizontal displacement is observed. This trend can be understood from the structure of Eq.~\ref{eq:langevin3}, where the wind velocity contributes additively to the droplet’s velocity obtained by solving Eq.~\ref{eq:langevin2}. As shown in Figure~\ref{fig:drop_radius_vs_Lmax2}, for droplet radii approaching $30\,\mu$m, the maximum horizontal distance can reach up to 150~m for Chlorpyrifos and nearly 140~m for HCB under a wind velocity of 10~m/s. 
\section{Conclusion and Outlook} 
\label{sec:conc}
In this study, we have explored the dispersion behavior of pesticide droplets under the influence of environmental factors such as wind, gravity, and fluid resistance, using a Langevin dynamics-based approach. Our formulation accounts for the stochastic evolution of droplet velocity and position, incorporating physical quantities such as the drag coefficient, fluid viscosity, and the temperature dependence of both density and diffusion. We assume that the diffusion process is modeled via a white noise term representing uncorrelated stochastic fluctuations. The wind is assumed to be constant across both horizontal and vertical directions, as well as over time. Additionally, we consider the spray release height to be fixed at an average person’s height of approximately 1.7 meters, and we assume a typical 5\% concentration strength of the chemical solution in the sprayed droplets. Seasonal variations, local weather fluctuations, diurnal changes, and other time-dependent environmental factors are not included in our current analysis.

We demonstrated how droplet properties—particularly size and chemical composition—affect the overall transport in air, and how wind speed dramatically alters the maximum horizontal distance a droplet can travel before settling. Our simulations reveal that smaller droplets are more susceptible to lateral drift due to wind, whereas larger droplets are dominated by gravitational settling~(Figures~\ref{fig:drop_radius_vs_Lmax} and~\ref{fig:drop_radius_vs_Lmax2}). This has significant implications for the efficiency of pesticide delivery, environmental contamination, and potential non-target exposure. The comparative analysis of different pesticide chemicals, including Chlorpyrifos and HCB, emphasizes the role of chemical-specific physical parameters in determining dispersion outcomes. Importantly, our study provides quantitative insight into the extent of pesticide drift under varying conditions. For example, under high wind speeds, droplets from pesticide-treated fields may travel distances of up to 120--150 meters. This finding underscores a regulatory concern—such droplet drift could potentially affect nearby lands designated for organic farming, thereby undermining their compliance with organic standards. Our results thus offer valuable input that may help revise or reinforce government-imposed buffer zone regulations between conventional and organic farmlands.

As a natural extension of this work, the simulation framework can be used to study the maximum permissible distance between conventionally treated farmland and designated organic farming zones. By incorporating seasonal variations and region-specific wind flow data across different parts of India, this analysis can help establish more accurate regulatory buffer zones to prevent pesticide drift into organic plots. This could be instrumental in refining existing government guidelines, especially considering that our simulations show potential lateral transport of pesticide droplets up to 130–150 meters under strong wind conditions, thereby posing a risk to nearby organic farms. Additionally, the framework can be adapted to scenarios involving aerial spraying by drones, where the initial spray height significantly influences droplet trajectory and dispersion. Such an extension would be valuable for understanding drift patterns in larger and more structured agricultural landscapes where drone-based application is increasingly adopted. Future developments may also incorporate complex effects such as droplet evaporation, non-spherical particle dynamics, and atmospheric turbulence, thereby enhancing the realism and applicability of the model.

\section*{Acknowledgement}
The authors acknowledge the funding support for project number $2015900$ from IBITF, IIT Bhilai, under which this research work was carried out. The authors thank Dr. Anindita Seth and Prof. Santosh Kumar Das for their valuable discussions and insightful suggestions, particularly during the early stages of formulating the problem.

\section*{DATA AVAILABILITY}
The data supporting the findings of this study are either included within the article or referenced appropriately.
%%%%%%%%%%%%%%%%%%%%%%%%%%%%%%%%%%%%%%%%%%%%%%%%%%%%%%%%%%%%%%%

\appendix
\section*{Appendix: Calculation of Pesticide Solution Density}
\label{app:A}
\addcontentsline{toc}{section}{Appendix: Calculation of Pesticide Solution Density}
To estimate the effective density of the insecticide solution used in our study, we consider a binary mixture of Chlorpyrifos (the solute) and water (the solvent). The overall density is calculated as a weighted average of the constituent densities, based on the mass fraction of the solute. Here we provide the details of the calculation for a 5\% solution at room temperature (\( T = 30\,^{\circ}\text{C} \)).

\begin{itemize}
  \item Concentration of insecticide solution: \( C = 5\% = 0.05 \)
  \item Temperature: \( T = 303\, \text{K} = 30\,^{\circ}\text{C} \)
  \item Solute density (Chlorpyrifos): \( \rho_{\text{solute}} = 1.40\, \text{g/cm}^3 \)
\end{itemize}

Using the empirical formula for the density of water in \(\text{g/cm}^3\)~\cite{doi:10.1021/je60064a005}, we have
\begin{align}
\rho_{\text{water}}(T) &= \frac{1}{1000} \left(999.842594 + 0.06793952~T - 0.009095290~T^2 + 0.0001001685~T^3 \right) \nonumber\\
\rho_{\text{water}}(30) &\approx 0.9957\, \text{g/cm}^3
\end{align}

Now, the effective density of the solution is a weighted average, expressed as:
\[
\rho_{\text{solution}} = C \cdot \rho_{\text{solute}} + (1 - C) \cdot \rho_{\text{water}}
\]

Substituting values from above, we get:
\begin{align}
\rho_{\text{solution}} &= 0.05 \cdot 1.40 + 0.95 \cdot 0.9957 \nonumber \\
&= 0.07 + 0.946915 \nonumber\\
\rho_{\text{solution}} &= 1.01692\, \text{g/cm}^3 = 1016.92\, \text{kg/m}^3
\end{align}

\bibliographystyle{utcaps_mod}
\bibliography{draft}

\providecommand{\href}[2]{#2}\begingroup\raggedright\begin{thebibliography}{10}

\bibitem{toxics11100858}
T.~Boonupara, P.~Udomkun, E.~Khan, and P.~Kajitvichyanukul, ``{\em Airborne
  Pesticides from Agricultural Practices: A Critical Review of Pathways,
  Influencing Factors, and Human Health Implications},''
  \href{http://dx.doi.org/10.3390/toxics11100858}{Toxics {\normalfont \bfseries
  11} (2023) no.~10, }. \url{https://www.mdpi.com/2305-6304/11/10/858}.

\bibitem{Socorro2016}
J.~Socorro, A.~Durand, B.~Temime-Roussel, S.~Gligorovski, H.~Wortham, and
  E.~Quivet, ``{\em The persistence of pesticides in atmospheric particulate
  phase: An emerging air quality issue},''
  \href{http://dx.doi.org/10.1038/srep33456}{Scientific Reports {\normalfont
  \bfseries 6} (2016)  33456}. \url{https://www.nature.com/articles/srep33456}.

\bibitem{vandenBerg1999}
F.~van~den Berg, R.~Kubiak, W.~G. Benjey, M.~S. Majewski, S.~R. Yates, G.~L.
  Reeves, J.~H. Smelt, and A.~M.~A. van~der Linden, ``{\em Emission of
  Pesticides into the Air},''
  \href{http://dx.doi.org/10.1023/A:1005234329622}{Water, Air, and Soil Pollution
  {\normalfont \bfseries 115} (1999)  195--218}.
  \url{https://link.springer.com/article/10.1023/A:1005234329622}.

\bibitem{Hofman2001}
V.~Hofman and E.~Solseng, ``{\em Reducing Spray Drift},'' tech. rep., North
  Dakota State University, June, 2001.
\newblock
  \url{https://library.ndsu.edu/ir/bitstream/handle/10365/9153/AE-1210.pdf}.

\bibitem{GIL20072945}
Y.~Gil, C.~Sinfort, Y.~Brunet, V.~Polveche, and B.~Bonicelli, ``{\em
  Atmospheric loss of pesticides above an artificial vineyard during
  air-assisted spraying},''
  \href{http://dx.doi.org/https://doi.org/10.1016/j.atmosenv.2006.12.019}{Atmospheric
  Environment {\normalfont \bfseries 41} (2007) no.~14, 2945--2957}.
  \url{https://www.sciencedirect.com/science/article/pii/S1352231006012647}.

\bibitem{Bengfort}
M.~Bengfort, H.~Malchow, and F.~Hilker,
  \href{http://dx.doi.org/10.1007/s00285-016-0966-8}{``{\em The Fokker–Planck
  law of diffusion and pattern formation in heterogeneous
  environments},''Journal of Mathematical Biology {\normalfont \bfseries 73}
  (09, 2016)  }.

\bibitem{Fick1855}
A.~Fick, ``{\em On Liquid Diffusion},'' Philosophical Magazine Journal of
  Science {\normalfont \bfseries 10} (1855)  31--39.

\bibitem{Fick1858}
A.~Fick, {\em Die medizinische Physik}.
\newblock Vieweg, Braunschweig, 1858.

\bibitem{Fokker}
A.~D. Fokker, ``{\em Die mittlere Energie rotierender elektrischer Dipole im
  Strahlungsfeld},''
  \href{http://dx.doi.org/https://doi.org/10.1002/andp.19143480507}{Annalen der
  Physik {\normalfont \bfseries 348} (1914) no.~5, 810--820},
  \href{http://arxiv.org/abs/https://onlinelibrary.wiley.com/doi/pdf/10.1002/andp.19143480507}{{\normalfont
  \ttfamily https://onlinelibrary.wiley.com/doi/pdf/10.1002/andp.19143480507}}.
  \url{https://onlinelibrary.wiley.com/doi/abs/10.1002/andp.19143480507}.

\bibitem{Planck1917}
M.~Planck, ``{\em {\"U}ber einen Satz der statistischen Dynamik und seine
  Erweiterung in der Quantentheorie},'' Sitzungsberichte der Preussischen
  Akademie der Wissenschaften zu Berlin {\normalfont \bfseries 24} (1917)
  324--341.

\bibitem{Dorr2013}
G.~Dorr, D.~H. Wasalathanthri, S.~I. Wyllie, B.~L. Keirnan, H.~Zhang, J.~W.~G.
  McLean, S.~J.~W. Spangenberg, T.~S. Langford, C.~J. Keatley, J.~R. O'Keeffe,
  R.~S. Franklin, T.~J. Watterson, and S.~W. Rowe, ``{\em Particle and droplet
  size analysis of pesticide spray drift from aerial applications},''
  \href{http://dx.doi.org/10.1038/nature12437}{Environmental Toxicology and
  Chemistry {\normalfont \bfseries 32} (2013) no.~10, 2294--2303}.
  \url{https://pubmed.ncbi.nlm.nih.gov/23945590/}.

\bibitem{Nuyttens}
D.~Nuyttens, M.~Schampheleire, P.~Verboven, E.~Brusselman, and D.~Dekeyser,
  \href{http://dx.doi.org/10.13031/2013.29127}{``{\em Droplet Size and Velocity
  Characteristics of Agricultural Sprays},''TRANSACTIONS OF THE ASABE
  {\normalfont \bfseries 52} (09, 2009)  1471--1480}.

\bibitem{van1999}
F.~van~den Berg, ``{\em Behavior of pesticides in air},''
  \href{http://dx.doi.org/10.1023/A:1005234329622}{Water, Air, and Soil
  Pollution {\normalfont \bfseries 115} (1999)  195--218}.

\bibitem{Koehler1995}
P.~G. Koehler and A.~H. Moye,
  \href{http://dx.doi.org/10.1093/jee/88.6.1684}{``{\em Airborne Insecticide
  Residues After Broadcast Application for Cat Flea (Siphonaptera: Pulicidae)
  Control},''Journal of Economic Entomology {\normalfont \bfseries 88} (12,
  1995)  1684--1689},
  \href{http://arxiv.org/abs/https://academic.oup.com/jee/article-pdf/88/6/1684/19238954/jee88-1684.pdf}{{\normalfont
  \ttfamily
  https://academic.oup.com/jee/article-pdf/88/6/1684/19238954/jee88-1684.pdf}}.
  \url{https://doi.org/10.1093/jee/88.6.1684}.

\bibitem{WU2014477}
X.~Wu, J.~C. Lam, C.~Xia, H.~Kang, Z.~Xie, and P.~K. Lam, ``{\em Atmospheric
  hexachlorobenzene determined during the third China arctic research
  expedition: Sources and environmental fate},''
  \href{http://dx.doi.org/https://doi.org/10.5094/APR.2014.056}{Atmospheric
  Pollution Research {\normalfont \bfseries 5} (2014) no.~3, 477--483}.
  \url{https://www.sciencedirect.com/science/article/pii/S1309104215303056}.

\bibitem{coulston1972}
F.~Coulston and F.~Korte, ``{\em Environmental quality and safety. Chemistry,
  toxicology and technology. Volume I. Global aspects of chemistry, toxicology
  and technology as applied to the environment.},''.

\bibitem{Koehler}
P.~G. Koehler and A.~H. Moye,
  \href{http://dx.doi.org/10.1093/jee/88.6.1684}{``{\em Airborne Insecticide
  Residues After Broadcast Application for Cat Flea (Siphonaptera: Pulicidae)
  Control},''Journal of Economic Entomology {\normalfont \bfseries 88} (12,
  1995)  1684--1689},
  \href{http://arxiv.org/abs/https://academic.oup.com/jee/article-pdf/88/6/1684/19238954/jee88-1684.pdf}{{\normalfont
  \ttfamily
  https://academic.oup.com/jee/article-pdf/88/6/1684/19238954/jee88-1684.pdf}}.
  \url{https://doi.org/10.1093/jee/88.6.1684}.

\bibitem{warner1980results}
S.~Warner, C.~Gerbig, R.~Strebing, and J.~Molello, ``{\em Results of a two-year
  toxicity and oncogenicity study of chlorpyrifos administered to CD-1 mice in
  the diet},'' Dow Chemical Co., Midland, MI (1980)  .

\bibitem{Schiott1970}
C.~R. Schiott, H.~L{\"o}e, S.~B. Jensen, M.~Kilian, R.~M. Davies, and
  K.~Glavind, ``{\em The effect of chlorhexidine mouthrinses on the human oral
  flora},'' \href{http://dx.doi.org/10.1111/j.1600-0765.1970.tb00697.x}{Journal
  of Periodontal Research {\normalfont \bfseries 5} (1970) no.~2, 84--89}.
  \url{https://pubmed.ncbi.nlm.nih.gov/4254173/}.

\bibitem{IARC1979Supplement1}
{International Agency for Research on Cancer}, {\em Chemicals and Industrial
  Processes Associated with Cancer in Humans (IARC Monographs Volumes 1 to
  20)}, vol.~1 of {\em IARC Monographs Supplement}.
\newblock IARC, Lyon, France, 1979.
\newblock
  \url{https://publications.iarc.fr/Book-And-Report-Series/Iarc-Monographs-Supplements/Chemicals-And-Industrial-Processes-Associated-With-Cancer-In-Humans-IARC-Monographs-Volumes-1-To-20--1979}.

\bibitem{IARC1979Vol20}
{International Agency for Research on Cancer}, {\em Some Halogenated
  Hydrocarbons}, vol.~20 of {\em IARC Monographs on the Evaluation of
  Carcinogenic Risks to Humans}.
\newblock IARC, Lyon, France, 1979.
\newblock
  \url{https://publications.iarc.fr/Book-And-Report-Series/Iarc-Monographs-On-The-Identification-Of-Carcinogenic-Hazards-To-Humans/Some-Halogenated-Hydrocarbons-1979}.

\bibitem{Cabral1986}
J.~R.~P. Cabral and P.~Shubik, ``{\em Carcinogenic activity of
  hexachlorobenzene in mice and hamsters},'' IARC Scientific Publications
  (1986) no.~77, 411--416. \url{https://pubmed.ncbi.nlm.nih.gov/3298036/}.

\bibitem{reif1965fundamentals}
F.~Reif, {\em Fundamentals of Statistical and Thermal Physics}.
\newblock Fundamentals of Physics Series. McGraw-Hill, 1965.
\newblock \url{https://books.google.co.in/books?id=3ApRAAAAMAAJ}.

\bibitem{Das2020}
S.~K. Das, J.~e~Alam, S.~Plumari, and V.~Greco,
  \href{http://dx.doi.org/10.1063/5.0022859}{``{\em Transmission of airborne
  virus through sneezed and coughed droplets},''Physics of Fluids {\normalfont
  \bfseries 32} (Sept., 2020)  097102}.

\bibitem{Teske2002AgDRIFT}
M.~E. Teske, S.~L. Bird, D.~M. Esterly, T.~B. Curbishley, S.~L. Ray, and S.~G.
  Perry, ``{\em {AgDRIFT: a model for estimating near-field spray drift from
  aerial applications}},''
  \href{http://dx.doi.org/10.1897/1551-5028(2002)021<0659:AAMFEN>2.0.CO;2}{Environmental
  Toxicology and Chemistry {\normalfont \bfseries 21} (2002) no.~3, 659--671}.

\bibitem{Merck1989}
S.~Budavari, ed., {\em The Merck Index: An Encyclopedia of Chemicals, Drugs,
  and Biologicals}.
\newblock Merck \& Co., Inc., Rahway, NJ, 11~ed., 1989.
\newblock Monograph No. 2194.

\bibitem{NTP_Hexachlorobenzene}
{National Toxicology Program}, {\em 15th Report on Carcinogens}.
\newblock U.S. Department of Health and Human Services, Public Health Service,
  2021.
\newblock \url{https://www.ncbi.nlm.nih.gov/books/NBK590945/}.

\bibitem{Verschueren1983}
K.~Verschueren, ``{\em Chlorpyrifos},'' in {\em Handbook of Environmental Data
  on Organic Chemicals}, K.~Verschueren, ed., p.~391.
\newblock Van Nostrand Reinhold Co., New York, NY, 2~ed., 1983.

\bibitem{EXTOXNET_HCB}
{Extension Toxicology Network}, ``{\em EXTOXNET Pesticide Information Profile:
  Hexachlorobenzene},'' 1998.
\newblock \url{https://extoxnet.orst.edu/pips/hexachlo.htm}.

\bibitem{white2006viscous}
F.~M. White and J.~Majdalani, {\em Viscous fluid flow}, vol.~3.
\newblock McGraw - Hill New York, 2006.

\bibitem{sutherland1893viscosity}
W.~Sutherland, ``{\em The Viscosity of Gases and Molecular Force},''
  \href{http://dx.doi.org/10.1080/14786449308620508}{The London, Edinburgh, and
  Dublin Philosophical Magazine and Journal of Science {\normalfont \bfseries
  36} (1893) no.~223, 507--531}.

\bibitem{stokes1851}
G.~G. Stokes, ``{\em On the effect of the internal friction of fluids on the
  motion of pendulums},'' Transactions of the Cambridge Philosophical Society
  {\normalfont \bfseries 9} (1851)  8--106.

\bibitem{beale1996statistical}
P.~D. Beale, {\em Statistical Mechanics}.
\newblock Butterworth-Heinemann, 1996.

\bibitem{doi:10.1021/je60064a005}
G.~S. Kell, ``{\em Density, thermal expansivity, and compressibility of liquid
  water from 0.deg. to 150.deg.. Correlations and tables for atmospheric
  pressure and saturation reviewed and expressed on 1968 temperature scale},''
  \href{http://dx.doi.org/10.1021/je60064a005}{Journal of Chemical \&
  Engineering Data {\normalfont \bfseries 20} (1975) no.~1, 97--105},
  \href{http://arxiv.org/abs/https://doi.org/10.1021/je60064a005}{{\normalfont
  \ttfamily https://doi.org/10.1021/je60064a005}}.
  \url{https://doi.org/10.1021/je60064a005}.

\end{thebibliography}\endgroup
\end{document}